\documentclass[letters,fleqn,usenatbib,useAMS]{mnras}

\usepackage{graphicx}	
\usepackage{amsmath}	

\newcommand{\bz}{\langle B_z \rangle}

\title[Magnetic fields in WR stars]{
The search for magnetic fields in two Wolf-Rayet stars and the discovery of a variable magnetic field in WR\,55 
}

\stepcounter{footnote}

\author[Hubrig et al.]{
S.~Hubrig$^{1}$\thanks{Corresponding author: shubrig@aip.de},
M.~Sch\"oller$^{2}$,
A.~Cikota$^{3}$,
S.~P.~J\"arvinen$^{1}$
\\
$^{1}${Leibniz-Institut f\"ur Astrophysik Potsdam (AIP), An der Sternwarte~16, 14482~Potsdam, Germany} \\
$^{2}${European Southern Observatory, Karl-Schwarzschild-Str.~2, 85748 Garching, Germany} \\
$^{3}${Physics Division, E.O. Lawrence Berkeley National Laboratory, 1
Cyclotron Road, Berkeley, CA 94720, USA} \\
}

\date{Accepted XXX. Received YYY; in original form ZZZ}

\pubyear{2019}

\begin{document}
\label{firstpage}
\pagerange{\pageref{firstpage}--\pageref{lastpage}}
\maketitle

\begin{abstract}
Magnetic fields in Wolf-Rayet (WR) stars are not well explored, although there is indirect evidence, e.g.\
from spectral variability and X-ray emission, that magnetic fields should be present in these stars.
Being in an advanced stage of their evolution, WR stars have lost their hydrogen 
envelope, but their dense winds make the stellar core almost unobservable.
To substantiate the expectations on the 
presence of magnetic fields in the most-evolved massive stars, we selected two WR stars,
WR\,46 and WR\,55, for the search of the presence of magnetic fields using FORS\,2 
spectropolarimetric observations.
We achieve a formally definite detection of a variable mean longitudinal magnetic field
of the order of a few hundred Gauss
in WR\,55. The field detection in this star, which is associated with the ring nebula 
RCW\,78 and the molecular environment, is of exceptional importance for our understanding of star formation.
No field detection at a significance level of 3$\sigma$
was achieved for WR\,46, but the variability of the measured field strengths can be rather well 
phased with the rotation period of 15.5\,h previously suggested by FUSE observations.
\end{abstract}

\begin{keywords}
  techniques: polarimetric --- 
  stars: individual: WR\,46 ---
  stars: individual: WR\,55 ---
  stars: magnetic fields ---
  stars: massive ---
  stars: Wolf-Rayet
\end{keywords}



\section{Introduction}
\label{sec:intro}

In the modeling of massive stars, specific aspects, especially the role of magnetic fields,
remain not well understood, implying large uncertainties in the 
star's evolutionary path and ultimate fate. 
Previous observations indicate that probably about 7\% of O-type stars with masses
exceeding 18\,${\rm M}_\odot$ and 
about 6\% of early B- and O-type stars have measurable,
mostly dipolar magnetic fields \citep[e.g.,][]{Grunhut2017,Schoeller2017}.
Theoretical models suggest that O stars with strong magnetic  
field detections may be related to magnetars with
$B\approx10^{15}$\,G \citep[e.g.][]{Thompson2004}.
Thus, it is important to 
detect magnetic fields in
WR stars, which are descendants of massive O stars and 
direct predecessors of compact remnants. 
WR stars are highly chemically evolved massive stars that have lost their hydrogen envelope and now expose their 
former stellar core. However, their
dense winds make the stellar surface almost unobservable.

Magnetic fields in WR stars are currently not sufficiently explored, in spite of the fact that
there is indirect evidence, e.g.\ from spectral variability and X-ray emission, that magnetic fields are
present in WR star atmospheres  \citep[e.g.][]{Michaux2014}. Previous theoretical work of 
\citet{Gayley2010} predicted a fractional
circular polarization of a few times 10$^{-4}$ for magnetic fields of about 100\,G. 
Using high-resolution spectropolarimetric observations with
ESPaDOnS at the Canada--France--Hawaii Telescope, \citet{Chevrotiere2014}, reported  
marginal magnetic field detections for WR\,134, WR\,137, and WR\,138, corresponding to magnetic field 
strengths
of about 200\,G, 130\,G, and\,80 G, respectively, and an average upper limit of about 500\,G for the 
non-detections in other stars.
As the line spectrum in WR stars is formed in the strong stellar wind, the detection of magnetic fields
in these stars is difficult.
The major problem is the wind broadening of the emission lines by Doppler shifts with wind velocities of 
a few thousand km\,s$^{-1}$. The broad spectral lines observed with high-resolution spectropolarimetry
extend over adjacent orders, so that it is necessary to adopt order shapes to get the best 
continuum normalization. 

Because of such immense line broadening in WR stars, to search for weak magnetic fields in a number 
of WR stars, \citet{Hubrig2016}
used the FOcal Reducer low dispersion Spectrograph in spectropolarimetric mode 
(FORS\,2; \citealt{Appenzeller1998}) mounted on the 8\,m Antu telescope of
the European Southern Obsevatory's
Very Large Telescope (VLT) on Cerro Paranal/Chile. 
The obtained FORS2 polarimetric spectra allowed \citet{Hubrig2016} to measure the mean longitudinal
magnetic field $\bz=258\pm78$\,G in
the cyclically variable and X-ray emitting WN5 star WR\,6 at a significance level of 3.3$\sigma$.
Keeping in mind that the two clearly magnetic Of?p stars HD\,148937 and CPD $-$28$^\circ$ 5104 were
for the first time detected as magnetic in the FORS\,2 observations at significance 
levels of 3.1$\sigma$ and 3.2$\sigma$, 
respectively \citep{Hubrig2008,Hubrig2011},
the detection of the magnetic field in WR\,6 at a significance level of 3.3$\sigma$ indicates that 
a magnetic field is likely present in this star.
Spectropolarimetric monitoring of WR\,6
revealed a sinusoidal nature of the $\left<B_{\rm z}\right>$ variations, which is indicative of a 
predominantly dipolar magnetic field structure (Hubrig et al.\ 2016; see Fig.~1, left side).
The field appeared to be reversing, with the extrema detected at rotation phases 0 and 0.5.

In the sample of spectropolarimetrically studied WR stars by \citet{Hubrig2016}, 
WR\,6 was the only target showing X-ray emission and 
cyclical variability due to the presence of corotating interacting regions (CIRs), which are formed
out of the interaction between high and low-velocity flows as the star rotates 
\citep[e.g.][]{Louis1995}.
CIRs were detected in spectroscopic time series observations of only a few
massive stars \citep[e.g.][]{Mullan1984}.  It was suggested that CIRs are related 
to the presence of magnetic bright spots, which are indicators of the presence of a global magnetic 
field \citep[e.g.][]{Rami2014}

To substantiate expectations on the 
presence of magnetic fields in the most-evolved massive stars, we have searched for magnetic 
fields in two other promising targets,  WR\,46 (=HD\,104994) and  WR\,55 (=HD\,117688), 
which, similar to WR\,6, show CIRs \citep[e.g.][]{Louis2011}.
These stars are accessible from the VLT and have never
been observed with spectropolarimetry in the past.

WR\,46 is a WN3p star \citep{Hamann2019}, with relatively strong \ion{O}{iv}~$\lambda$~3811 and $\lambda$~3834
emission lines. It is very hot and 
compact ($T_{\rm eff}=112.2$\,kK, $R=1.4\,{\rm R}_\odot$), and
also bright in X-rays,
possessing a hard component in its X-ray emission \citep{Gosset2011}, which may indicate 
the magnetic nature of WR\,46. 
According to \citet{Brunet2011}, WR\,46 is known to exhibit a very complex variability 
pattern.  The different periods and timescales observed in the past suggest the presence of multiple periods,
including dominant and secondary periods (see Fig.~1 in their work).
To explain the short-term variability of this star, various scenarios were evoked, including
the possibility of a close binary or non-radial pulsations \citep[e.g.][]{Veen2002a,Veen2002b}.
Using observations with the Far Ultraviolet Spectroscopic Explorer (FUSE), \citet{Brunet2011} found significant 
variations on a timescale of $\sim$8\,h. This period is close to the photometric and spectroscopic periods
previously reported by other authors.  \citet{Brunet2011} also reported the 
detection of a second significant peak, just slightly weaker, corresponding to 
$P=15.5\pm2.5$\,h.  

WR\,55 is a significantly cooler ($T_{\rm eff}=56.2$\,kK) WN7 star with hydrogen deficiency and 
belongs to the WNE subclass, like WR\,6 \citep{Hamann2006}.
However, its radius, $R=5.2\,{\rm R}_\odot$, is larger compared to the radius of WR\,6 with $R=3.2\,{\rm R}_\odot$.
A highly significant level of spectroscopic variability
of about 10\% was discovered by \citet{Louis2011}. So far,
no periodicity search was carried out for this star. \citet{Cappa2009}
investigated the distribution 
of molecular gas related to the ring nebula RCW\,78 around WR\,55 and concluded
that WR\,55 is not only responsible for the ionization of the gas in the
nebula, but also for the creation of the interstellar bubble. Their analysis indicates that the star
formation in this region is induced by the strong wind of this star. Thus, the discovery of a magnetic field
in WR\,55 would be of exceptional interest for star formation theories.

As WR stars are characterized by
spectra showing very broad emission lines, the determination of their magnetic fields is usually based on the
calculation of the mean longitudinal magnetic field, i.e.\ of the line-of-sight field component, using
circularly polarized light.
To search for magnetic fields in WR\,46 and WR\,55, we obtained several randomly timed distributed low-resolution
spectropolarimetric observations using
the FOcal Reducer low dispersion Spectrograph \citep[FORS\,2;][]{Appenzeller1998},
installed at the ESO/VLT. 
In the following, we give an overview of our spectropolarimetric
observations, describe the data reduction and discuss
the results of the magnetic field measurements. 


\section{Data reduction and results of the magnetic field measurements}
\label{sect:obs}

To maximize the field detection probability and to avoid missing the magnetic field due
to an unfavorable viewing angle in certain rotation phases,
FORS\,2 observations of both targets were obtained on several
different epochs to sample different rotation phases.
Seven spectropolarimetric observations of WR\,46 were obtained in service mode, one observation
on 2016 March~14 and six observations
from 2020 January~16 to February~21. For WR\,55, we obtained five observations from 2020 February~13 to March~10.
The last observation of  WR\,55 recorded on March 10 was not completed, probably due to 
bad weather conditions, and was therefore used only for the inspection of spectral variability.

The FORS\,2 multi-mode instrument is equipped with polarisation analysing optics
comprising super-achromatic half-wave and quarter-wave phase retarder plates,
and a Wollaston prism with a beam divergence of 22$\arcsec$ in standard
resolution mode. 
We used the GRISM 600B and the narrowest available slit width
of 0$\farcs$4 to obtain a spectral resolving power of $R\sim2000$
in the observed spectral range from 3250 to 6215\,\AA{}.
For the observations, we used a non-standard readout mode with low 
gain (200kHz,1$\times$1,low), which provides a broader dynamic range, hence 
allowed us to reach a higher signal-to-noise ratio ($S/N$) in the individual spectra.
The circular polarization observations were carried out using a
sequence of positions of the quarter-wave plate $-45^{\circ}$, $+45^{\circ}$, $+45^{\circ}$, $-45^{\circ}$
and so forth, 
to minimize the cross-talk effect and to cancel errors from different
transmission properties of the two polarized beams. Moreover, the
reversal of the quarter-wave plate compensates for fixed errors in
the relative wavelength calibrations of the two polarized spectra.
The ordinary and extraordinary beams
were extracted using standard IRAF procedures as described by \citet{Cikota2017}.
The wavelength calibration was carried out using He-Ne-Ar arc lamp exposures.

\begin{figure}
\centering 
\includegraphics[width=0.45\textwidth]{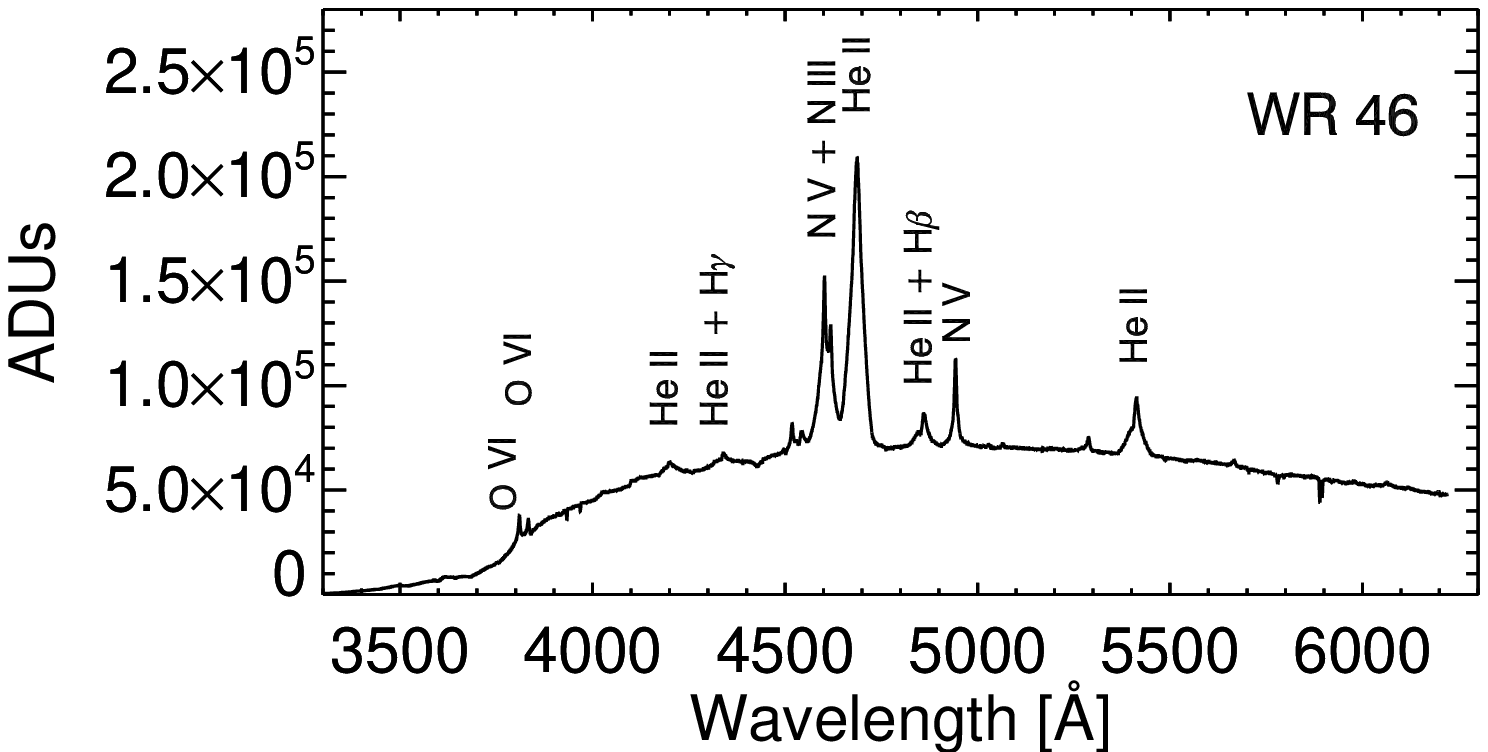}
\includegraphics[width=0.45\textwidth]{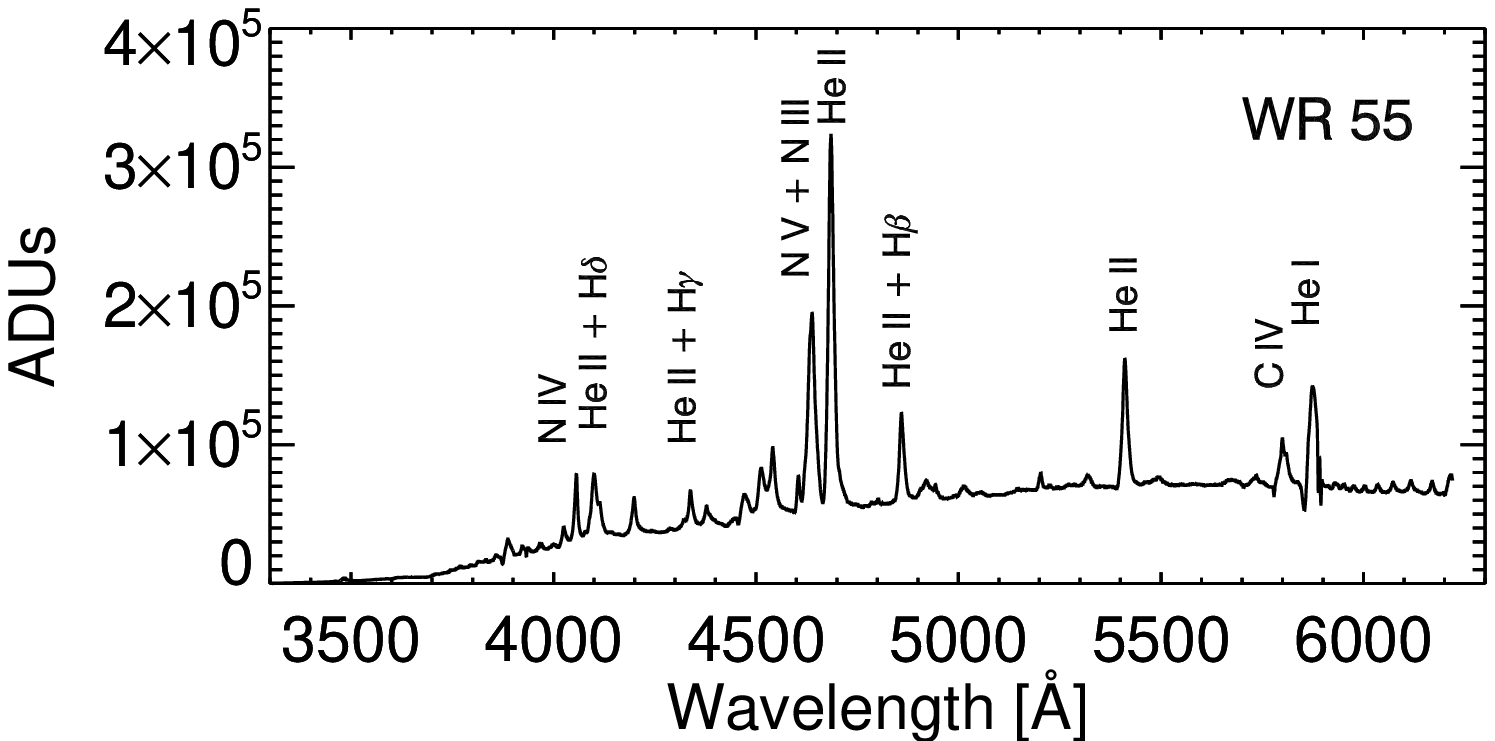}
        \caption{
FORS 2 Stokes~$I$ spectra of  WR\,46 and WR\,55.
Spectral line identification is based on the 
work of \citet{Hamann2006} and is shown above the line profiles.
}
   \label{fig:sp}
\end{figure}

The spectral appearance of the WN stars WR\,46 and WR\,55 in the FORS 2 spectra is presented in Fig.~\ref{fig:sp}.
The spectra of WN stars are dominated by helium and nitrogen lines.
The WN3p star WR\,46 is characterized by the presence of strong \ion{N}{v} and \ion{He}{ii} lines 
and the absence of hydrogen. The “p” stands for peculiar and denotes the presence of unusually  
strong \ion{O}{vi}~$\lambda$~3811 and $\lambda$~3834 emission lines, the relatively strong
\ion{N}{v}~$\lambda$~4604 line,
and relatively weak \ion{C}{iv}~$\lambda$~5801 and ~$\lambda$~5812 lines (e.g.\ \citealt{Conti1989}).

\begin{figure}
 \centering 
\includegraphics[width=0.43\textwidth]{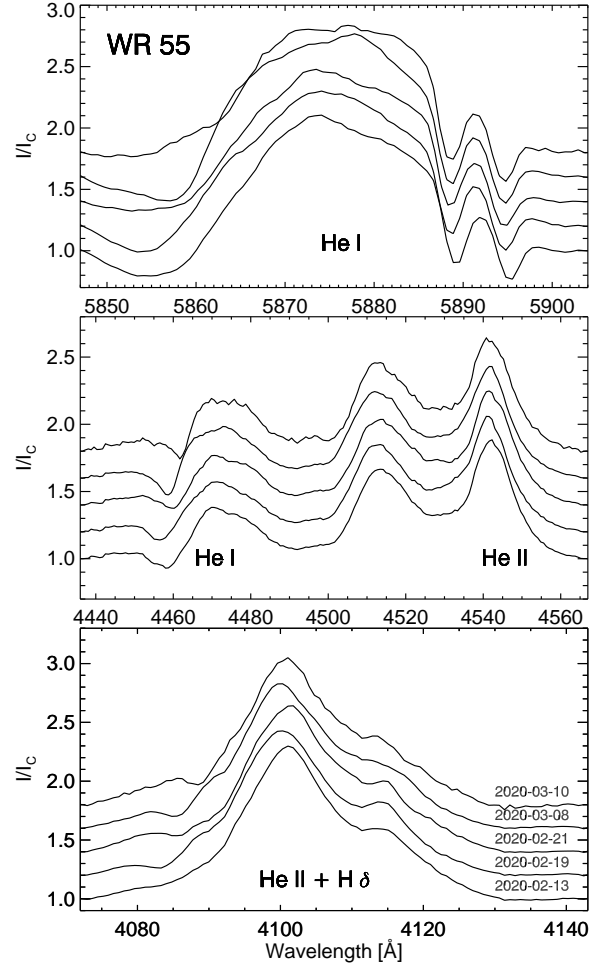}        
        \caption{
FORS\,2 Stokes~$I$ spectra showing variability of different spectral lines.
The spectra obtained at different epochs are offset vertically for better visibility.
Spectral line  identification is based on the work of \citet{Hamann2006}.}
   \label{fig:mos3}
\end{figure}

While previous observations of WR\,46 clearly showed the presence of variability in photometry and spectroscopy,
the variability of WR\,55 was studied only once by \citet{Louis2011} using
 spectra in the spectral range 5200--6000\,\AA{} obtained with the 1.5\,m
telescope of the Cerro Tololo Inter-American Observatory. 
The authors reported for this star a highly significant level of variability of up to 10\% of the 
line intensity. As we show in Fig.~\ref{fig:mos3}, spectral line variability  
is also detected in our FORS\,2 Stokes $I$ spectra of this star.

A description of the assessment of the presence of a longitudinal magnetic
field using FORS\,1/2 spectropolarimetric observations was presented in our
previous work
\citep[e.g.][and references therein]{Hubrig2004a,Hubrig2004b}.
Improvements to the methods used, including $V/I$
spectral  rectification  and  clipping,  were  detailed  by  \citet{Hubrig2014}.
Null spectra are calculated as pairwise differences from all available
$V$ profiles so that the real polarisation signal should cancel out. From
these, 3$\sigma$-outliers are identified and used to clip the $V$ profiles.
This removes spurious signals, which mostly come from cosmic rays, and also
reduces the noise.

The mean longitudinal magnetic field $\left< B_{\rm z}\right>$
is measured on the rectified and clipped spectra,
based on the relation following the method suggested by
\citet{Angel1970}:

\begin{eqnarray}
\frac{V}{I} = -\frac{g_{\rm eff}\, e \,\lambda^2}{4\pi\,m_{\rm e}\,c^2}\,
\frac{1}{I}\,\frac{{\rm d}I}{{\rm d}\lambda} \left<B_{\rm z}\right>\, ,
\label{eqn:vi}
\end{eqnarray}

\noindent
where $V$ is the Stokes parameter that measures the circular polarization,
$I$ is the intensity in the unpolarized spectrum, $g_{\rm eff}$ is the effective
Land\'e factor, $e$ is the electron charge, $\lambda$ is the wavelength,
$m_{\rm e}$ is the electron mass, $c$ is the speed of light,
${{\rm d}I/{\rm d}\lambda}$ is the wavelength derivative of Stokes~$I$, and
$\left<B_{\rm z}\right>$ is the mean longitudinal (line-of-sight) magnetic
field.

\begin{table}
\caption{Longitudinal magnetic field values obtained for WR\,46 and WR\,55 using
  FORS\,2 observations. In the first column we show the modified Julian dates
  of mid-exposures, followed by the corresponding signal-to-noise ratio 
($S/N$) of the FORS\,2 Stokes~$I$ spectra measured close to 4686\,\AA{}. The measurements 
of the mean longitudinal magnetic field using the
  Monte Carlo bootstrapping test and using the null spectra are presented in Columns~3 and 4.
All quoted errors are 1$\sigma$ uncertainties. 
}
\label{tab:meas}
\centering
\begin{tabular}{lrr@{$\pm$}lr@{$\pm$}l}
\hline
\multicolumn{1}{c}{MJD} &
\multicolumn{1}{c}{$S/N$} &
\multicolumn{2}{c}{$\left<B_{\rm z}\right>$} &
\multicolumn{2}{c}{$\left<B_{\rm z}\right>_N$} \\
\multicolumn{1}{c}{} &
\multicolumn{1}{c}{} &
\multicolumn{2}{c}{(G)} &
\multicolumn{2}{c}{(G)} \\
\hline
\multicolumn{6}{c}{WR\,46} \\
\hline
57461.3213  & 2339 &$-$199 & 88 &    46 &  81 \\
58864.2960  & 1477 &  342 & 154 & $-$68 & 146 \\
58885.1674  & 1845 &   35 &  94 & $-$62 & 109 \\
58885.2627  & 2046 &$-$43 &  99 & $-$7  &  97 \\
58892.1191  & 1189 &  268 & 194 &   81  & 173 \\
58898.3843  & 1905 &   23 &  93 & $-$41 & 115 \\
58900.1569  & 1088 &$-$112& 160 &    23 & 179 \\
\hline
\multicolumn{6}{c}{WR\,55} \\
\hline
58892.2058 & 2388 &   205& 58 & $-$16 & 55 \\
58898.3492 & 2579 &$-$378& 85 &  $-$2 & 78 \\
58900.2075 & 2168 &    56& 80 &    63 & 82 \\
58916.3059 & 2386 &     4& 66 & $-$16 & 57 \\
58918.2531 & 533  &     \multicolumn{2}{c}{}    &    \multicolumn{2}{c}{}   \\
\hline
\end{tabular}
\end{table}

Furthermore, we have carried out Monte Carlo bootstrapping tests. These are
most often applied with the purpose of deriving robust estimates of standard
errors (e.g.\ \citealt{Steffen2014}). 
The measurement uncertainties obtained before and after the Monte Carlo
bootstrapping tests were found to be in close agreement, indicating the
absence of reduction flaws. The results of our magnetic field measurements
are presented in Table~\ref{tab:meas}.

\begin{figure}
 \centering 
        \includegraphics[width=0.45\textwidth]{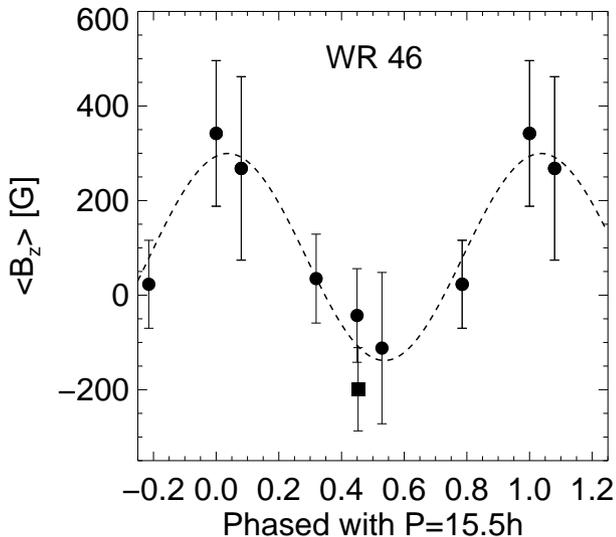}
        \caption{
Longitudinal magnetic field measurements of WR\,46 carried out using low-resolution FORS\,2  
spectropolarimetric observations and phased with the period of 15.5\,h. The overplotted dashed curve 
corresponds  to the sinusoidal  fit. The filled square corresponds to the observation obtained in 2016.
}
   \label{fig:mos1}
\end{figure}

For WR\,46, the values for the longitudinal magnetic field
$\bz$ show change of polarity,
with the strongest mean longitudinal magnetic field
of positive polarity $\bz=342\pm154$\,G at a significance
level of 2.2$\sigma$ and the strongest field of negative polarity $\bz=-199\pm88$\,G at a significance
level of 2.3$\sigma$.
Since significant photometric and spectroscopic 
variations on a timescale of $\sim$8\,h were reported in previous studies of this star,
assuming that this periodicity is caused by rotational modulation, we tested 
the distribution of the measurement values over this period. We do not find any 
hint for sinusoidal modulation, which is expected for a large-scale organized dipole field structure.
However, as we show in Fig.~\ref{fig:mos1}, rotation modulation  is indicated in our data, if we use 
the period of 15.5\,h suggested by \citet{Brunet2011}. Due to the large uncertainty of this period,
only the measurements obtained in 2020 are fitted by a sinusoid. 
The older measurement obtained 
in 2016 and marked by the filled square appears (purely coincidentally)
slightly shifted from the expected negative field 
extremum. Interestingly, our period analysis by fitting a sinusoid to the measurements obtained in 2020,
indicates almost the same rotation period $P_{\rm rot}=15.53\pm0.14$\,h.
Obviously, the fidelity of the 15.5\,h period needs to be confirmed with a
long-term extensive data set.

For WR\,55, the values for the longitudinal magnetic field
$\bz$ show change of polarity 
with the highest field values 
$\bz=-378\pm85$\,G at a significance
level of 4.4$\sigma$ and $\bz=205\pm58$\,G at a significance
level of 3.5$\sigma$.
In view of the importance of the field detection at a significance
level of 4.4$\sigma$, we decided to carry out a consistency check using a
different spectral extraction.
The parallel and perpendicular beams in the observations at this epoch 
were extracted using a pipeline written in the MIDAS environment and developed by
T. Szeifert, the very first FORS instrument scientist. 
More details on this pipeline can be found in \citet{Hubrig2014}.
The result of this measurement, $\bz=-334\pm77$\,G,
is fully compatible with the measurement $\bz=-378\pm85$\,G within the error bars.

The simplest model for the magnetic field geometry in stars with globally organized fields is based on the 
assumption that the studied stars are oblique dipole rotators, i.e.\ their magnetic
field can be approximated by a dipole with the magnetic axis inclined to the
rotation axis.
Unfortunately, the rotation axis inclination $i$ for WR stars is undefined
because of their dense winds, making the measurement of the projected rotation velocity
$v\,\sin\,i$ using broad emission lines impossible.
Since the rotation period and the limb-darkening are also unknown for WR\,55,
we can only estimate for this star a minimum dipole strength of $\sim$1.13\,kG using the relation
$B_{\rm d} \ge 3 \left| \left<B_{\rm z}\right>_{\rm max} \right|$ \citep{Babcock1958}.


\section{Discussion}
\label{sec:meas}

Although magnetic fields are now believed to play an important role in the
evolution of massive stars, spectropolarimetric observations of WR stars, which are descendants
of massive O stars, are still very scarce.
WR stars are usually rather faint, and, in addition, their line spectra are formed in the 
strong stellar wind, with the wind broadening of the emission lines up to a few thousand km\,s$^{-1}$.
Both WR\,46 and WR\,55 are faint with visual magnitudes $m_{\rm v}\ge10.9$, and have never been observed 
spectropolarimetrically in the past. So far, the presented FORS\,2 observations are
the first to explore the magnetic nature of these targets.
The strongest mean longitudinal magnetic field for WR\,46 was measured at a 
significance level of 2.2$\sigma$ at the positive field extremum and at the level 2.3$\sigma$ 
close to the negative field extremum.
Only for WR\,55 we achieve formally definite detections, $\left<B_{\rm z} \right> = -378\pm85$\,G at a 
significance level of 4.4$\sigma$ and $\left<B_{\rm z} \right> = 205\pm58$\,G at a significance level 
of 3.5$\sigma$.
We should, however, keep in mind that the magnetic
field is diagnosed in the line-forming regions, which fall fairly far
out in a WR wind, and not at the stellar surface. Different lines form over different zones of the wind, 
sampling different field strengths. Thus, the  method we apply for the measurements
only gives a result for the field strength where lines are formed. It is expected that the surface
values of the magnetic field are significantly stronger than the field measured in the wind lines 
\citep{chevro2013}.

With respect to the confidence of the field detection, a recent detailed comparison
between our analysis technique and an independent analysis from another team 
showed that the measurement 
results  agree  well within expected  statistical  distributions \citep{Schoeller2017}. This  gives
us high confidence about the accuracy of our longitudinal magnetic
field measurements. Importantly, detections at a $\le3$~$\sigma$ level appear to be 
genuine in a number of studies where the measurements show smooth variations over a rotation period,
similar to those found for the magnetic
Of?p stars HD 148937 and  CPD\,$-$28$^{\circ}$\,2561 \citep{Hubrig2008,Hubrig2011,Hubrig2013,Hubrig2015}.
In these studies, not a 
single reported detection reached a 4$\sigma$ significance level.

The first detection of the presence of a magnetic field in WR\,55 makes this star the
best candidate for long-term spectropolarimetric monitoring. Future observing 
campaigns should be based on spectropolarimetric time series to further strengthen the evidence for the 
magnetic nature of this star and to set more stringent limits to its  magnetic
field strength. The temporal variations
of the measured longitudinal magnetic fields should be used to ascertain the rotation/magnetic period
of WR\,55 and determine for the first time the geometry of the global magnetic field in a WR star.

Furthermore, the detection of the magnetic field in WR\,55 associated with the ring nebula RCW\,78 and its molecular 
environment is of an exceptional importance for our understanding of star formation. 
According to \citet{Cappa2009}, 
WR\,55 is not only responsible for the ionization of the gas in the
nebula, but also for the creation of the interstellar bubble. The  presence of star formation activity  
in the environment of this nebula suggests that it may have been triggered by the expansion
of the bubble.


\section*{Acknowledgements}
Based on observations made with ESO Telescopes at the La Silla Paranal
Observatory under the programme IDs 097.D-0428(A) and 0104.D-0246(A).
SPJ is supported by the German Leibniz-Gemeinschaft, project number P67-2018.
We thank the referee G. Mathys for his constructive comments.
\section*{Data Availability}

The FORS\,2 data from 2016 are available from the ESO Science Archive Facility at
http://archive.eso.org/cms.html.
The data from 2020 will become available in March 2021 at the same location
and can be requested from the author before that date.

\bsp	
\label{lastpage}

\begin{thebibliography}{99}

\bibitem[\protect\citeauthoryear{Angel \& Landstreet}{1970}]{Angel1970}
Angel J.~R.~P., Landstreet J.~D.,
1970, ApJ, 160, L147

\bibitem[\protect\citeauthoryear{Appenzeller et al.}{1998}]{Appenzeller1998} 
Appenzeller I., et al.,
1998, The ESO Messenger, 94, 1

\bibitem[\protect\citeauthoryear{Babcock}{1958}]{Babcock1958}
Babcock H.~W., 1958, ApJS, 3, 141

\bibitem[\protect\citeauthoryear{Cappa et al.}{2009}]{Cappa2009} 
Cappa C.~E., Rubio M., Mart\'in M.~C., Romero G.~A.,
2009, A\&A, 508, 759

\bibitem[\protect\citeauthoryear{Chen\'e \& St-Louis}{2011}]{Louis2011}
Chen\'e A.-N., St-Louis N.,
2011, ApJ, 736, 140

\bibitem[\protect\citeauthoryear{Cikota et al.}{2017}]{Cikota2017} 
Cikota A., Patat F., Cikota S., Faran T.,
2017, MNRAS, 464, 4146

\bibitem[\protect\citeauthoryear{Conti \& Massey}{1989}]{Conti1989}
Conti P.~S., Massey P.,
1989, ApJ, 337, 251 

\bibitem[\protect\citeauthoryear{de la Chevroti\'ere et al.}{2013}]{chevro2013}
de la Chevroti\'ere A., St-Louis N., Moffat A.~F.~J., MiMeS Collaboration,
2013, ApJ, 764, 171 

\bibitem[\protect\citeauthoryear{de la Chevroti\'ere et al.}{2014}]{Chevrotiere2014}  
de la Chevroti\'ere A., St-Louis N., Moffat A.~F.~J., MiMeS Collaboration,
2014, ApJ, 781, 73 

\bibitem[\protect\citeauthoryear{Gayley \& Ignace}{2010}]{Gayley2010}
Gayley K.~G., Ignace R.,
2010, ApJ, 708, 615

\bibitem[\protect\citeauthoryear{Gosset et al.}{2011}]{Gosset2011}  
Gosset E., De Becker M., Naz\'e Y., Carpano S., Rauw G., Antokhin I.~I., Vreux J.-M., Pollock A.~M.~T.,
2011, A\&A, 527, A66

\bibitem[\protect\citeauthoryear{Grunhut et al.}{2017}]{Grunhut2017}
Grunhut J.~H., et al.,
2017, \mnras, 465, 243

\bibitem[\protect\citeauthoryear{Hamann et al.}{2006}]{Hamann2006} 
Hamann W.-R., Gr\"afener G., Liermann A.,
2006, A\&A, 457, 1015

\bibitem[\protect\citeauthoryear{Hamann et al.}{2019}]{Hamann2019} 
Hamann W.-R., et al.,
2019, A\&A, 625, A57 

\bibitem[\protect\citeauthoryear{H\'enault-Brunet et al.}{2011}]{Brunet2011}
H\'enault-Brunet V., St-Louis N., Marchenko S.~V., Pollock A.~M.~T., Carpano S., Talavera A.,
2011, ApJ, 735, 13

\bibitem[\protect\citeauthoryear{Hubrig et al.}{2004a}]{Hubrig2004a}
Hubrig S., Kurtz D.~W., Bagnulo S., Szeifert T., Sch\"oller M., Mathys G., Dziembowski W.~A.,
2004a, A\&A, 415, 661

\bibitem[\protect\citeauthoryear{Hubrig et al.}{2004b}]{Hubrig2004b}
Hubrig S., Szeifert T., Sch\"oller M., Mathys G., Kurtz D. W.,
2004b, A\&A, 415, 68

\bibitem[\protect\citeauthoryear{Hubrig et al.}{2008}]{Hubrig2008}
Hubrig S., Sch\"oller M., Schnerr R.~S., Gonz\'alez J.~F., Ignace R., Henrichs H.~F.,  
2008, A\&A, 490, 793

\bibitem[\protect\citeauthoryear{Hubrig et al.}{2011}]{Hubrig2011}
Hubrig S., et al.,
2011, A\&A, 528, A151

\bibitem[\protect\citeauthoryear{Hubrig et al.}{2013}]{Hubrig2013}
Hubrig S., et al.,
2013, A\&A, 551, A33

\bibitem[\protect\citeauthoryear{Hubrig et al.}{2014}]{Hubrig2014}
Hubrig S., Sch\"oller M., Kholtygin A. F., 2014,
MNRAS, 440, L6

\bibitem[\protect\citeauthoryear{Hubrig et al.}{2015}]{Hubrig2015}
Hubrig S., et al.,
2015, MNRAS, 447, 1885

\bibitem[\protect\citeauthoryear{Hubrig et al.}{2016}]{Hubrig2016}
Hubrig S., et al., 2016,
MNRAS, 458, 3381

\bibitem[\protect\citeauthoryear{Michaux et al.}{2014}]{Michaux2014}
Michaux Y.~J.~L., Moffat A.~F.~J., Chen\'e A.-N., St-Louis N.,
2014, MNRAS, 440, 2

\bibitem[\protect\citeauthoryear{Mullan}{1984}]{Mullan1984}
Mullan D.~J.,
1984, ApJ, 283, 303

\bibitem[\protect\citeauthoryear{Ramiaramanantsoa et al.}{2014}]{Rami2014}
Ramiaramanantsoa T., et al.,
2014, MNRAS,  441, 910

\bibitem[\protect\citeauthoryear{Sch\"oller et al.}{2017}]{Schoeller2017}
Sch\"oller M., et al.,
2017, \aap, 599, A66

\bibitem[\protect\citeauthoryear{St-Louis et al.}{1995}]{Louis1995}
St-Louis N., Dalton M.~J., Marchenko S.~V., Moffat A.~F.~J., Willis A.~J.,
1995, ApJ, 452, L57

\bibitem[\protect\citeauthoryear{Steffen et al.}{2014}]{Steffen2014}
Steffen M., Hubrig S., Todt H., Sch\"oller M., Hamann W.-R., Sandin C., Sch\"onberner D.,
2014, A\&A, 570, A88

\bibitem[\protect\citeauthoryear{Thompson et al.}{2004}]{Thompson2004}
Thompson T.~A., Chang P., Quataert E.,
2004, ApJ, 611, 380

\bibitem[\protect\citeauthoryear{Veen et al.}{2002a}]{Veen2002a}
Veen P.~M., Van Genderen A.~M., Crowther P.~A., van der Hucht K.~A. 2002a, A\&A, 385, 600

\bibitem[\protect\citeauthoryear{Veen et al.}{2002b}]{Veen2002b}
Veen P.~M., Van Genderen A.~M., van der Hucht K.~A. 2002b, A\&A,~385, 619

\end{thebibliography}
\end{document}